\documentclass[pdflatex,sn-mathphys-num]{sn-jnl}
\usepackage{amsmath,amssymb,amsfonts}
\usepackage{mathtools} 
\usepackage{graphicx}
\usepackage{xcolor}
\hypersetup{colorlinks=true,pdfborder={0 0 0.1},linkcolor=orange!50!black,urlcolor=blue!30!black,citecolor=blue!60!black,
          pdftitle={Tensor cross interpolation for global discrete optimization with application to Bayesian network inference},
          pdfauthor={Sergey Dolgov, Dmitry Savostyanov},
          }

\usepackage[ruled]{algorithm}
\usepackage{algorithmic}
\def\eps{\varepsilon}
\def\geq{\geqslant} \def\leq{\leqslant}
\def\Real{\mathbb{R}} \def\Binary{\{0,1\}} 

\def\X{\mathbb{X}}            

\def\inf{\text{(inf)}} \def\rec{\text{(rec)}}   

\def\phi{\varphi}
\def\Prob{\mathsf{P}}   \def\Like{\mathsf{L}} 
\def\emptyset{\varnothing}  

\def\new{\star}
\def\trans{T}
\def\x{\mathrm{x}} \def\y{\mathrm{y}} \def\e{\mathrm{e}}    
\def\g{\mathbf{\hat g}}  
\def\I{\mathcal{I}}
\def\Rnd{\mathcal{R}}
\def\new{\star}

\def\Nodes{\mathcal{V}} \def\Edges{\mathcal{E}}  
\def\Grid{\mathcal{G}} \def\Grids{\mathbb{G}}
\def\Data{\mathcal{X}}

\def\neval{N_{\mathrm{eval}}}
\def\tmax{T} 
\def\tstep{\Delta t}
\def\Temp{\tau}

\begin{document}

\title[Tensor cross interpolation for global discrete optimization]{Tensor cross interpolation for global discrete optimization with application to Bayesian network inference\footnote{%
 S.D. is supported by the Engineering and Physical Sciences Research Council New Investigator Award EP/T031255/1.}}

\author*[1]{\fnm{Sergey} \sur{Dolgov}}\email{S.Dolgov@bath.ac.uk}
\equalcont{These authors contributed equally to this work.}

\author[2]{\fnm{Dmitry} \sur{Savostyanov}}\email{D.Savostyanov@essex.ac.uk}
\equalcont{These authors contributed equally to this work.}

\affil*[1]{\orgname{University of Bath}, \orgaddress{\street{Claverton Down}, \city{Bath}, \postcode{BA2 7AY}, \country{UK}}}
\affil[2]{\orgname{University of Essex}, \orgaddress{\street{Wivenhoe Park}, \city{Colchester}, \postcode{CO4 3SQ}, \country{UK}}}

\abstract{
Global discrete optimization is notoriously difficult due to the lack of gradient information and the curse of dimensionality, making exhaustive search infeasible.
Tensor cross approximation is an efficient technique to approximate multivariate tensors (and discretized functions) by tensor product decompositions based on a small number of tensor elements, evaluated on adaptively selected fibers of the tensor, that intersect on submatrices of (nearly) maximum volume.
The submatrices of maximum volume are empirically known to contain large elements, hence the entries selected for cross interpolation can also be good candidates for the globally maximal element within the tensor.
In this paper we consider evolution of epidemics on networks, and infer the contact network from observations of network nodal states over time.
By numerical experiments we demonstrate that the contact network can be inferred accurately by finding the global maximum of the likelihood using tensor cross interpolation. 
The proposed tensor product approach is flexible and can be applied to global discrete optimization for other problems, e.g. discrete hyperparameter tuning.
}

\maketitle

\section{Introduction}  \label{sec:intro}

Global optimization of functions that may have multiple local optima is an infamously challenging problem, particularly in high dimensions. 
Continuous global optimization is more researched, and many well--performing methods exist, such as
  genetic algorithms~\cite{tulu-stga-2004,venkatraman-ga-2005,weise-2009},
  consensus--based optimization~\cite{hajinkim-cbo-2021,totzeck-cbo-2022,Kalise-CO-2023,pareschi-cbo-2023,riedl-cbo-saddle-2024,riedl-cbo-global-2024}, or
  Markov Chain Monte Carlo~\cite{jejeva-mcmcopt-2020,jejebe-mcmcopt-2021}.
In contrast, discrete global optimization of multivariate functions is short of state of the art methods, for the following two main reasons: 
  \begin{enumerate}
   \item the absence of gradients --- and more generally the lack of continuity in the discrete setting, which makes many efficient methods from the list above impossible to apply;
   \item the curse of dimensionality --- the search space size grows exponentially with the number of variables, hence a brute force search is unfeasible in more than a dozen of dimensions.
  \end{enumerate}
Methods based on the Markov process framework (now over a discrete state set) are one of the few feasible approaches.
However, without gradient enhancements they may converge slowly and get stuck in local optima frequently.

A novel and promising approach to multivariate optimization, and generally to solving problems in high dimensions, is based on tensor product approximations.
To introduce the concept, consider a function $f(\x)$ defined on a Cartesian product of $N$ discrete sets 
\(
\X \times \cdots \times \X =: \X^N \ni \x.
\)
The function values $f(\x)$ can be collected into a $N$-dimensional array 
\(
\begin{bmatrix} f(\x) \end{bmatrix}_{\x\in\X^N},
\)
also called a \emph{tensor}.
The tensor has $(\#\X)^N$ entries, making it unfeasible to compute or store for large $N,$ which is known as the curse of dimensionality.
However, if the tensor admits a low--rank approximation, it may be possible to compute it efficiently and work with this representation instead.

In this paper we use the Tensor-Train (TT) approximation~\cite{osel-tt-2011}, noting that other tensor product formats and algorithms are available, see e.g.~\cite{hackbusch-2012}.
Even if the tensor itself is large and inaccessible in full, its TT approximation can be computed efficiently by tensor cross interpolation~\cite{ot-ttcross-2010,sav-qott-2014,ds-parcross-2020} from a number of adaptively selected samples, which shape one--dimensional lines, or \emph{fibers}, in the given tensor.
A fiber is a subset of positions 
\(
 \x^{(s)} = (x_1,\ldots,x_{n-1},x_n^{(s)},x_{n+1},\ldots,x_N),
\)
\(
s=1,\ldots,(\#\X)
\)
in the tensor, such that only one index $x_n$ is varying, and the other $N-1$ indices are frozen. 
This approach is inspired by alternating least square optimization algorithms used in statistics~\cite{harshman-parafac-1970,cc-parafac-1970,tuckerals-1980,zhang-mals-2003} and quantum physics~\cite{schollwock-2011,ds-amen-2014,schollwock-dmrg3s-2015}.
On each step, positions of the next fiber(s) are chosen adaptively in order to maximize the volume (absolute value of the determinant)~\cite{gtz-maxvol-1997,gostz-maxvol-2010} of the matrix of tensor elements formed by the intersection in each cross.
Maximum--volume submatrices are empirically known to contain large elements, hence the entries selected for cross interpolation can also be good candidates for the globally maximal element within the tensor.
Hence, a natural idea is to approximate a given tensor using tensor cross interpolation algorithm, and look for the maximum (in modulus) value of tensor among the entries chosen by the maximum--volume principle.
Crucially, the number of samples per iteration is linear in $N$ and $\#\X$, hence the outlined procedure is computationally feasible and allows a direct search for the maximum among the evaluated samples.
The concrete way to generate fibers may vary.
One option is to apply the so-called \emph{maxvol} iteration algorithm~\cite{gostz-maxvol-2010} to reshaped factors (\emph{cores}) of the TT decomposition directly~\cite{ot-ttcross-2010}.
Optimization based on this method was used successfully to speed up
   the search for molecular docking configurations~\cite{ZheOfeKat-ttdock-2013,Sulimov-docking-2017,sulimov-dock-2019,Zheltkov-ttopt-proc-2020,Zheltkov-ttopt-2020}
   and reinforcement learning~\cite{scspco-ttopt-2022}.

In this paper we apply a slightly different tensor cross interpolation with a greedy search for fibers~\cite{sav-qott-2014,ds-parcross-2020} to global discrete optimization which arises in Bayesian inference of a contact network from the observed dynamics of epidemic evolving on this network. 
The problem boils down to finding the network with the largest a posteriori estimate, which in absence of prior knowledge is equivalent to maximization of the likelihood function.
Evaluating likelihood for each contact network is rather numerically expensive, since it requires 
  either solving a high--dimensional system of ODEs, known as Markovian master equation, or 
  accurately estimating the likelihood by extensive application of the Gillespie's stochastic simulation algorithm (SSA)~\cite{gillespie-ssa-1976}.
Thus, we prefer the greedy cross interpolation instead of the \emph{maxvol} TT cross algorithm~\cite{ot-ttcross-2010}, because in our numerical experience the former tends to require fewer samples than the latter.

Although the main contribution of this paper consists in testing the efficiency of the existing method applied to a relatively novel problem, we also suggest, test and evaluate a number of technical improvements to the greedy tensor cross interpolation algorithm~\cite{sav-qott-2014}, namely tempering and caching. 
Tempering introduces a parameter called temperature, which modifies the likelihood function and changes the high--dimensional landscape for interpolation and optimization.
Caching simply means keeping all evaluated elements of the tensor, even those that are tested but not chosen by the adaptive tensor cross interpolation algorithm, as they may be tested again (and even selected)  at later steps of the method. 
We demonstrate numerically that both enhancements considerably speed up the global optimization of the likelihood.

The rest of the paper is organised as follows.
In Sec.~\ref{sec:bayes} we describe the general framework for the Bayesian inference of the contact network from epidemiological data.
In Sec.~\ref{sec:esis} we present the specific epidemiological model, denoted $\eps$-SIS, that we use for the experiments in this paper.
In Sec.~\ref{sec:cross} we recall the tensor cross interpolation algorithm from~\cite{sav-qott-2014,ds-parcross-2020} and state it using notation related to the specific problem considered in this paper to simplify the presentation.
In Sec.~\ref{sec:tempering} we describe tempering and caching.
In Sec.~\ref{sec:num} we apply the proposed method to Bayesian network inference of a simple contact network. We demonstrate that the tensor cross interpolation can be efficient as a heuristic algorithm for global discrete optimization, and that caching and tempering additionally improve its performance.

\newpage
\section{Bayesian network identification model} \label{sec:bayes}
Following from our previous work~\cite{ds-infer-2024}, 
 we consider the problem of identifying the network of contacts between nodes (species, individuals, groups, etc.)
 from observed discrete states of these nodes at given time points of a continuous--time Markov process.
The network is an unweighted undirected graph
 \(
 \Grid = (\Nodes,\Edges),
 \)
 consisting of a set of nodes
 \(
 \Nodes = \{1,2,\ldots,N\}
 \)
 and links (or edges)
 \(
 \Edges = \{(m,n):\: m\in\Nodes, n\in\Nodes, \: m\neq n\},
 \)
 such that
 \(
  (m,n) \in\Edges \:\Leftrightarrow\: (n,m)\in\Edges.
 \)
 We will say that nodes $m,n \in \Nodes$ are connected,
 \(
 m\sim n,
 \)
 iff
 \(
 (m,n)\in\Edges.
 \)
Each node can take one of discrete states,
 \(
 x_n \in \X,
 \)
e.g. susceptible, infected, recovered.
The state of the whole network is therefore a vector
\(
 \x = \begin{pmatrix}x_1 & x_2 & \ldots & x_N \end{pmatrix}^\trans \in \X^N.
\)

We consider a continuous--time Markov jump process on the state space $\Omega=\X^N,$
assuming that the probability to switch from a state $\x$ to a state $\y$ in a infinitesimal time interval $dt$ is $p_{\x\to\y}^{\Grid}dt$
with some known transition rate $p_{\x\to\y}^{\Grid}$
that depends on the connectivity of the given network $\Grid$.
For example, the probability of infection spreading between nodes $n$ and $m$ can be nonzero only if $m \sim n$.
The system dynamics is described by the probabilities of network states
\(
p(\x,t|\Grid) = \Prob(\text{system on a network $\Grid$ is in state $\x$ at time $t$}).
\)
This probability function satisfies a system of ordinary differential equations (ODEs),
known as \emph{Markovian master equation}~\cite{vankampen-stochastic-1981,chen-stoch-sir-2005}, \emph{Chapman} or \emph{forward Kolmogorov equations}:
\begin{equation}\label{eq:cme}
 p'(\x,t|\Grid)
  = \sum_{\y\in\X^N}
        \left(
          p_{\y\to\x}^{\Grid} \cdot p(\y,t|\Grid) - p_{\x\to\y}^{\Grid}\cdot p(\x,t|\Grid)
        \right), \qquad \x\in\X^N.
\end{equation}

In practice, we can measure only particular states at particular points in time, which we collect into a dataset
\(
\Data = \{t_k,\x(t_k)\}_{k=0}^K
\)
of samples of the random variable $X(t)$ with the probability distribution function $p(\x,t|\Grid)$.
We seek the maximum likelihood estimator (MLE) of the network configuration,
\begin{equation}\label{eq:loglike}
 \begin{split}
  \Grid_{\text{opt}} & = \arg\max_{\Grid\in\Grids}\Like(\Grid),
 \end{split}
\end{equation}
where the likelihood of the dataset $\Data$ is defined as
\begin{equation}\label{eq:like}
 \begin{split}
  \Like(\Grid)
          & = \Prob(\Data | \Grid)
            = \Prob( X(t_1)=\x_1, \cdots, X(t_K)=\x_K | \Grid),
 \end{split}
\end{equation}
and $\Grids$ is some prior set of admissible networks, for example, all networks with unweighted edges.
If a uniform prior is used, 
\(
\Grid_{\text{opt}}
\)
is also the maximum a posteriori (MAP) estimator.
Since the system dynamics is Markovian, the likelihood can be expanded as follows
\begin{equation}\label{eq:markov}
 \begin{split}
  \Like(\Grid)
          & = \Prob( X(t_1)=\x_1 | X(t_0)=\x_0, \Grid) \cdots \Prob( X(t_K)=\x_K | X(t_{K-1})=\x_{K-1}, \Grid),
     \\   & = \prod_{k=1}^K \underbrace{\Prob( X(t_k)=\x_k | X(t_{k-1})=\x_{k-1}, \Grid)}_{\Prob(\x_{k-1}\to \x_k | \Grid)}.
 \end{split}
\end{equation}
The latter conditional probability is the solution to the master equation~\eqref{eq:cme} started from $100\%$ probability assigned to $\x_{k-1}$ as the initial state, and integrated over $[t_{k-1}, t_k]$,
\begin{equation}\label{eq:cmelike}
\Prob(\x_{k-1}\to \x_k | \Grid)  = p(\x_k, t_k | \Grid), \quad \text{given} \quad p(\x, t_{k-1} | \Grid) = \begin{cases}1, & \x = \x_{k-1}, \\ 0, & \text{otherwise.}\end{cases}
\end{equation}
The master equation~\eqref{eq:cme} can be solved 
  by stochastic simulation algorithms~\cite{gillespie-ssa-1976,Gillespie2001,hemberg-perfect-sampling-2007,AndersonHigham-MLCME-2012,Yates-MLCME-2016},
  or directly, approximating the probability function $p(\x,t|\Grid)$ for example by sparse grids~\cite{hegland-cme-2007},
  adaptive finite state projections~\cite{munsky-fsp-2006,jahnke-wavelet-cme-2010,Cao-FSP-2016},
  radial basis functions~\cite{Schuette-RBF-CME-2015},
  neural networks~\cite{Khammash-NN-CME-2021,Grima-CME-NN-2022},
  and tensor product approximations, such as
   canonical polyadic (CP) format~\cite{jahnke-cme-2008,Ammar-cme-2011,hegland-cme-2011},
   and TT format~\cite{kkns-cme-2014,dkh-cme-2014,ds-amen-2014,d-tamen-2018,Sidje-TT-CME-2017,Dinh-QTT-CME-2020,Ion-TT-CME-2021,Schuette-CME-CO-2016,ds-ttsir-2024}.
In this paper we use the latter approach~\cite{d-tamen-2018,ds-ttsir-2024} since the likelihoods require both 
  breaking the curse of dimensionality for large $N$, and
  estimating even small probabilities accurately, such that the error is below $10^{-3}$--$10^{-4}$.

\section{$\varepsilon$-SIS model process and network parametrization} \label{sec:esis}
\begin{figure}[t]
  \includegraphics[width=\linewidth]{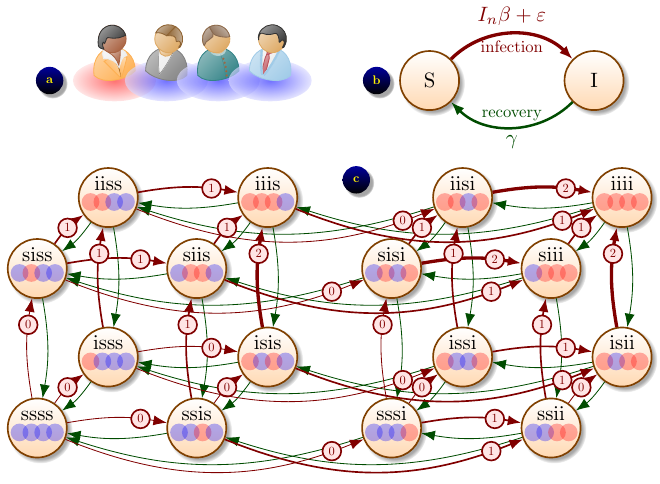}
 \caption{
  Markov chain and transitions between network states for the $\varepsilon$-SIS infection process. 
       (a) In this example, the contact network is a chain of $N=4$ people.
       (b) Transition rates in the Markov process depend not just on the size of infected and susceptible group, but on the exact state of the network. Note that the infection rate for node $n$ is $I_n \beta + \varepsilon$, where 
    $I_n$ is the number of infected neighbours of node $n$,
    $\beta$ is per contact infection rate, and
    $\varepsilon$ is the self--infection rate, responsible for contacts with potential threats outside the network.
    (c) All network states, with recovery transitions shown with green arrows, and infection transition shown with red arrows with a circled number $k$ indicating the infection rate $k\beta+\varepsilon.$
 }
 \label{fig:esis}
\end{figure}
As a particular model, we consider the so-called $\varepsilon$-SIS process, as shown in Fig.~\ref{fig:esis}.
Each node can take only one of two states:
 \(
 x_n \in \X
  = \{\text{susceptible}, \text{infected}\}
  = \{0,1\},
  \)
i.e. there is no recovered state. 
The system exhibits only two types of transitions:
\begin{equation}\label{eq:reactions}
 p_{\x\to\y}^{\Grid} =
  \begin{cases}
   p_{\x\to\y}^\inf, &\text{if $\exists n\in\Nodes,$ such that $\y=\x+\e_n$} \qquad \text{(infection);} \\
   p_{\x\to\y}^\rec, &\text{if $\exists n\in\Nodes,$ such that $\y=\x-\e_n$} \qquad \text{(recovery);} \\
    0,           &\text{otherwise,}
  \end{cases}
\end{equation}
where $\e_n\in\Real^N$ is the $n$--th unit vector.
For the simplicity sake, recovery rates $p_{\x\to\y}^\rec=\gamma$ are considered to be the same for all infected nodes.
The infection rate for a node $n$ (nonzero only if $n$ is susceptible, i.e. $x_n=0$) is set to
 \(
 p_{\x\to\y}^\inf=I_n(\x)\beta+\eps.
 \)
Note that it grows with the number of infected neighbouring nodes,
 \(
 I_n(\x) = \#\{ m\in\Nodes: m\sim n, x_m=1 \} ,
 \)
at the constant per-contact rate $\beta,$
on top of the baseline infection rate $\eps$ introduced from outside the network.

With known number of nodes and constant rates $\beta,\gamma,\varepsilon$ of the only possible transitions in the $\varepsilon$--SIS process, the last bit of information comes from a contact network $\Grid.$ 
The unweighted network can be represented by adjacency matrix of edges, $G \in \Binary^{n \times n},$ where $g_{m,n}=1$ indicates that $m$ and $n$ are connected, and $g_{m,n}=0$ means they are not.
Since the graph $\Grid$ is undirected, its adjacency matrix is symmetric, $G=G^\trans.$
Without loss of generality we take $g_{n,n}=0.$
Under these assumptions the graph $\Grid$ is defined uniquely by the elements of the adjacency matrix $G$ above the diagonal.
 We will collect the $\tfrac12 N(N-1)$ elements from the upper triangular part of $G$ and stretch them in a single \emph{adjacency vector} $\g\in\Binary^d,$ where $d:=\tfrac12 N(N-1),$ and the elements are arranged as follows 
\begin{equation}
  \g = \begin{pmatrix}\hat g_1 & \ldots & \hat g_{d} \end{pmatrix}^\trans 
    := \begin{pmatrix}g_{1,2} & g_{1,3} & g_{2,3} & \ldots & g_{N-1,N} \end{pmatrix}^\trans \in \Binary^d.
\end{equation}
Since $\Grid \Leftrightarrow \g,$ the likelihood~\eqref{eq:like} can also be seen as a function of the adjacency vector,
\(
\Like(\Grid) = \Like(\g): \Binary^d \to \Real,
\)
and the array of all likelihoods 
\(
 \begin{bmatrix} \Like(\Grid) \end{bmatrix}_{\Grid\in\Grids}
\)
can be seen as a $2\times 2\times \cdots \times 2$ tensor with $d$ binary indices.

\newpage
\section{Tensor Train cross approximation and optimization} \label{sec:cross}
All possible likelihood values $\Like(\g)$ for all possible combinations of $\g$ can be collected into a tensor of order $d$, with elements
\(
\Like(\hat g_1,\ldots,\hat g_{d}) = \Like(\g).
\)
In this section we consider low--rank cross approximation methods for approximation and optimization of this tensor.

\subsection{Matrix cross interpolation}
Consider first only two variables $\g = (\hat g_1, \hat g_2)$, such that the likelihood tensor is a matrix with elements $\Like(\hat g_1, \hat g_2)$.
To introduce a nontrivial low--rank matrix approximation (which will be necessary in higher-dimensional case), let us relax the assumption $\g \in \Binary^2$ for the moment, and consider $\g \in \{1,\ldots,\hat M\} \times \{1,\ldots,\hat N\}.$
If the matrix $\Like\in\Real^{\hat M \times \hat N}$ has rank $r \leq\min(\hat M, \hat N)$, it can be interpolated exactly via any $r$ of its linearly independent rows and columns,
 \begin{equation}\label{eq:mat}
  \begin{split}
  \Like(\hat g_1, \hat g_2) = \tilde \Like(\hat g_1, \hat g_2)
         & := \sum_{s=1}^r \sum_{t=1}^r \Like(\hat g_1,\I_2^{(t)}) [\Like(\I_1,\I_2)]_{t,s}^{-1} \Like(\I_1^{(s)},\hat g_2)
      \\ & = \Like(\hat g_1,\I_2) [\Like(\I_1,\I_2)]^{-1} \Like(\I_1,\hat g_2),
  \end{split}
 \end{equation}
where $\I_1 \subset \{1,\ldots,\hat M\}$ is a set of row indices and $\I_2 \subset \{1,\ldots,\hat N\}$ is a set of column indices.
In general, one can aim at approximating $\tilde \Like \approx \Like$.
The approximation \eqref{eq:mat} is still exact on the positions of the computed rows $\I_1$ and columns $\I_2$, which together form a cross of elements of $\Like$, hence the name \emph{cross interpolation}.
However, to minimize the error in the other elements
it becomes crucial to select $\I_1,\I_2$ that draw not just linearly independent, but as well-conditioned as possible rows and columns.
Theoretical guarantees are available if $\I_1,\I_2$ are chosen such that the intersection matrix has maximum \emph{volume},
\(
|\det \Like(\I_1,\I_2)| \to \max
\)
over all possible $\I_1 \subset \{1,\ldots,\hat M\}$ and $\I_2 \subset \{1,\ldots,\hat N\}$ of size $r$~\cite{gtz-maxvol-1997,gtz-psa-1997,gostz-maxvol-2010,schneider-cross2d-2010,gt-skel-2011}.
By replacing the inverse of $\Like(\I_1,\I_2)$ by the Moore--Penrose pseudoinverse,
the results have been generalized to rectangular submatrices~\cite{zo-maxvol-2017,mo-rectmaxvol-2018}.
However, to guarantee that a $r \times r$ submatrix of a $\hat M \times \hat N$ matrix with \emph{both} $\hat M,\hat N > r$ has the maximum volume is NP--hard~\cite{bartholdi-1982}.
In practice, cheaper heuristic algorithms are used.
For example, the \emph{incomplete cross approximation}~\cite{tee-cross-2000,ot-ttcross-2010} is based on fixing e.g. $\I_2$ to some initial guess, and searching only $\I_1$ as a maximum--volume set of rows in a smaller $\hat M \times r$ matrix, and vice versa.
This so-called \emph{maxvol} algorithm works well for many approximation and optimization problems; see e.g.~\cite{ro-hf-2016,dafs-tt-bayes-2019,dc-dirt-2020,dks-gradient-cross-2023,ado-ttrisk-2023,ZheOfeKat-ttdock-2013,Zheltkov-ttopt-2020,scspco-ttopt-2022} for a very incomplete list of examples.
However, for our needs it is somewhat inconvenient: the corresponding interpolation formula interpolates only after the full iteration over all variables is completed, which may require quite a few of expensive samples from $\Like$.

\begin{figure}[t]
  \begin{center}
  \includegraphics[width=.8\linewidth]{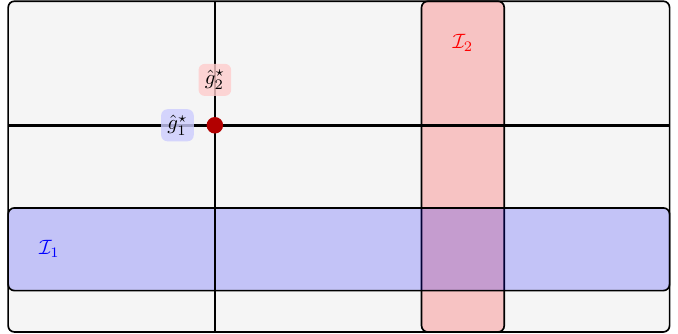}
  \end{center}
  \caption{One step of the matrix cross interpolation algorithm} \label{fig:mat}
\end{figure}

The greedy algorithm for the matrix cross interpolation that we use in this paper can be seen as a version of the \emph{adaptive cross approximation} (ACA) algorithm~\cite{bebe-2000}, or as the Gaussian elimination with partial pivoting~\cite{golub-2013}.
The algorithm builds up the subsets of rows and columns for the interpolation by subsequently adding one row and column, which is characteristic for \emph{greedy} algorithms.
Specifically, we use the \emph{rook} pivoting~\cite{poole-1992}, which searches for a leading index $(\hat g_1^\new,\hat g_2^\new)$ in both row and column of the residual spawned from $\hat g_1^\new$ and $\hat g_2^\new$,
\begin{equation}\label{eq:rook}
 \left|\Like(\hat g_1^\new,\hat g_2^\new)-\tilde \Like(\hat g_1^\new,\hat g_2^\new)\right| 
     \geq \left|\Like(\hat g_1,\hat g_2)-\tilde \Like(\hat g_1,\hat g_2)\right|, 
\end{equation}
for all $(\hat g_1,\hat g_2)$ such that $\hat g_1=\hat g_1^\new$ or $\hat g_2=\hat g_2^\new.$
The rook pivoting still needs an initial guess for $\hat g_1^\new,\hat g_2^\new$.
We search for one among $\min(\hat N,\hat M)$ indices $\hat g_1,\hat g_2$ sampled uniformly at random, as explained in Alg.~\ref{alg:mat} and visualised in Fig.~\ref{fig:mat}.
Note that one iteration of Alg.~\ref{alg:mat} needs at most $\hat M+\hat N$ evaluations of $\Like$, and both the approximation error and candidate maxima can be updated frequently.

\begin{algorithm}[t]
\caption{One step of the matrix cross interpolation algorithm} \label{alg:mat}
\begin{algorithmic}[1]
\REQUIRE Current row and column sets $\I_1,\I_2$.
\STATE Pick a random set of samples $\Rnd=\{(\hat g_1,\hat g_2)\}$ and choose the one with the largest error,
	\par
	$
	(\hat g_1^\new,\hat g_2^\new) \gets \arg\max_{(\hat g_1,\hat g_2)\in\Rnd} |\Like(\hat g_1,\hat g_2) - \tilde \Like(\hat g_1,\hat g_2)|
	$
\REPEAT
	\STATE
	$
	\hat g_1^\new \gets \arg\max_{\hat g_1\in\{1,\ldots,\hat M\}\backslash \I_1} |\Like(\hat g_1,\hat g_2^\new) - \tilde \Like(\hat g_1,\hat g_2^\new)|
	$
	\STATE
	$
	\hat g_2^\new \gets \arg\max_{\hat g_2\in\{1,\ldots,\hat N\}\backslash \I_2} |\Like(\hat g_1^\new,\hat g_2) - \tilde \Like(\hat g_1^\new,\hat g_2)|
	$
\UNTIL{rook condition~\eqref{eq:rook} holds \textbf{or} maximal number of iterations reached}
\ENSURE Expanded sets $\I_1 \gets \I_1 \cup \{\hat g_1^\new\}$, $\I_2 \gets \I_2 \cup \{\hat g_2^\new\},$ maximal error $\mathcal{E} = |\Like(\hat g_1^\new,\hat g_2^\new) - \tilde \Like(\hat g_1^\new,\hat g_2^\new)|.$
\end{algorithmic}
\end{algorithm}

\subsection{TT cross interpolation algorithm}
Matrix methods can be applied to tensors reshaped to matrices.
We introduce \emph{multi--indices}
\[
\hat g_{\leq k}=\hat g_1\hat g_2\ldots \hat g_{k}, \qquad
\hat g_{>k}=\hat g_{k+1}\ldots \hat g_d, \quad \forall k=1,\ldots,d-1.
\]
Those allow us to reshape a tensor, e.g. 
\(
\Like \in \Real^{2 \times \cdots \times 2},
\)
to a matrix 
\(
\Like^{\{k\}} \in \Real^{2^k \times 2^{d-k}},
\)
with
\[
\Like^{\{k\}}(\hat g_{\leq k},\hat g_{>k}) = \Like^{\{k\}}(\hat g_1\hat g_2\ldots \hat g_{k};\hat g_{k+1}\ldots \hat g_d) = \Like(\hat g_1,\hat g_2,\ldots,\hat g_d),
\]
The matrix interpolation formula~\eqref{eq:mat} can be written for $\Like^{\{k\}},$
\begin{equation}\label{eq:k}
 \begin{split}
 \Like^{\{k\}}(\hat g_{\leq k},\hat g_{>k})
    & \approx \tilde \Like^{\{k\}}(\hat g_{\leq k},\hat g_{>k})
 \\ & = \Like^{\{k\}}(\hat g_{\leq k},\I_{>k}) [\Like^{\{k\}}(\I_{\leq k},\I_{> k})]^{-1} \Like^{\{k\}}(\I_{\leq k},\hat g_{>k}).
 \end{split}
\end{equation}
However, since the ranges of $\hat g_{\leq k}$ and $\hat g_{>k}$ still suffer from the curse of dimensionality for intermediate $k$, direct usage of \eqref{eq:k} is not practical.

Nonetheless, one of the factors of \eqref{eq:k} is feasible for borderline $k$.
For example, for $k=1$, we can assume that some small enough index set $\I_{>1}$ is given.
Without a better initial guess, we can initialise it at random.
The left factor $\Like^{\{1\}}(\hat g_{\leq 1},\I_{>1}) = \Like(\hat g_1, \I_{>1})$
is sufficiently small, and can be assembled.
Similarly, the left index set $\I_{\leq 1}$ can be assembled and/or updated by the information in $\Like(\hat g_1, \I_{>1})$.
This sets the base for a recursive factorization of 
\(
\Like(\I_{\leq 1}, \hat g_2, \ldots, \hat g_d).
\)

\begin{algorithm}[t]
  \caption{Left--to--right sweep of the greedy TT cross interpolation algorithm~\cite{sav-qott-2014}} \label{alg:cross-iter}
  \begin{algorithmic}[1]
  \REQUIRE Sets $(\I_{\leq k},\I_{>k})$ of the interpolation~\eqref{eq:tt}
  \FOR{$k=1,\ldots,d-1$}
    \STATE Apply Alg.~\ref{alg:mat} to the subtensor $\Like(\I_{\leq k-1}^\new \hat g_k,\hat g_{k+1}\I_{>k+1})$ seen as a $(r_{k-1}\cdot 2 ) \times (2 \cdot r_{k+1})$ matrix. 
    Find
     the new pivot $(\hat g_{\leq k}^\new,\hat g_{>k}^\new)$ and
     the maximal error $\mathcal{E}_k = |\Like(\hat g_{\leq k}^\new,\hat g_{>k}^\new) - \tilde\Like(\hat g_{\leq k}^\new,\hat g_{>k}^\new)|$
    \STATE $\I_{\leq k}^\new  \gets \I_{\leq k}\cup\{\hat g_{\leq k}^\new\};$
          $\I_{>k}^\new \gets \I_{>k}\cup\{\hat g_{>k}^\new\}$
  \ENDFOR
  \ENSURE Updated index sets $(\I_{\leq k},\I_{>k}) \gets (\I_{\leq k}^\new,\I_{>k}^\new),$ $k=1,\ldots,d-1,$ maximal error $\mathcal{E} = \max_{k} \mathcal{E}_k.$
  \end{algorithmic}
\end{algorithm}

For a general $k \in \{1,\ldots,d-1\}$, we can assume that both $\I_{\leq k-1} = \{\hat g_{\leq k-1}\}$ of size $r_{k-1}$ and $\I_{>k+1}=\{\hat g_{>k}\}$ of size $r_{k+1}$ are available.
For $k=1$ or $k=d$ we can let $\I_{\leq 0} = \I_{>d} = \emptyset$.
The corresponding subtensor defined by 
\(
\Like(\I_{\leq k-1}, \hat g_k, \hat g_{k+1}, \I_{>k+1})
\)
contains only $4 r_{k-1} r_{k+1}$ elements.
We can reshape it into a $2r_{k-1} \times 2 r_{k+1}$ matrix with elements 
\(
\Like(\I_{\leq k-1}\hat g_k; \hat g_{k+1}\I_{>k+1}),
\)
and factorize it using the matrix cross interpolation algorithm.
Moreover, if some $\I_{\leq k}$ and $\I_{>k}$ are already available too,
we don't need to assemble $\Like(\I_{\leq k-1}\hat g_k; \hat g_{k+1}\I_{>k+1})$ in full,
but we can apply Algorithm~\ref{alg:mat} starting with $\I_{\leq k}$ and $\I_{>k}$ to find only one new pivot at a time.
This gives us a cross interpolation
\begin{equation}\label{eq:supercore}
 \begin{split}
  \Like(\I_{\leq k-1}\hat g_k; \hat g_{k+1}\I_{>k+1}) 
     &\approx \tilde\Like(\I_{\leq k-1}\hat g_k; \hat g_{k+1}\I_{>k+1}) 
   \\& := \Like(\I_{\leq k-1}, \hat g_k, \I_{>k})  [\Like(\I_{\leq k},\I_{>k})]^{-1} \Like(\I_{\leq k}, \hat g_{k+1}, \I_{>k+1}).
 \end{split}
\end{equation}
Recall that in the base of the recursion we left behind the left factor $\Like(\hat g_1, \I_{>1}) [\Like(\I_{\leq 1}, \I_{>1})]^{-1}$.
Completing the recursion, we arrive at the following high--dimensional cross interpolation:
\begin{equation}\label{eq:tt}
 \begin{split}
 \Like(\hat g_1,\ldots,\hat g_d)
   & \approx \tilde \Like(\hat g_1,\ldots,\hat g_d)\\
   & := \Like(\hat g_1,\I_{>1}) [\Like(\I_{\leq 1},\I_{>1})]^{-1} \Like(\I_{\leq 1},\hat g_2,\I_{>2}) [\Like(\I_{\leq 2},\I_{>2})]^{-1}  \cdots  \Like(\I_{\leq d-1},\hat g_d).
 \end{split}
\end{equation}
Merging the inverse matrices with either left or right factors, we see that~\eqref{eq:tt} resembles the traditional TT decomposition~\cite{osel-tt-2011}
\begin{equation}\label{eq:tt0}
 \tilde\Like(\hat g_1,\ldots,\hat g_d) 
   = \Like^{(1)}(1,\hat g_1,:) \Like^{(2)}(:,\hat g_2,:) \cdots \Like^{(d)}(:, \hat g_d,1),
\end{equation}
where $\Like^{(k)} \in \Real^{r_{k-1} \times 2 \times r_k}$ are the \emph{TT cores},
such that $\Like^{(k)}(:,\hat g_k,:) \in \Real^{r_{k-1} \times r_k}$ for $k=1,\ldots,d$, with $r_0=r_d=1$,
and the sizes of TT cores and index sets $r_1,\ldots,r_{d-1}$ are called \emph{TT ranks}.

\begin{figure}[t]
  \includegraphics[width=\linewidth]{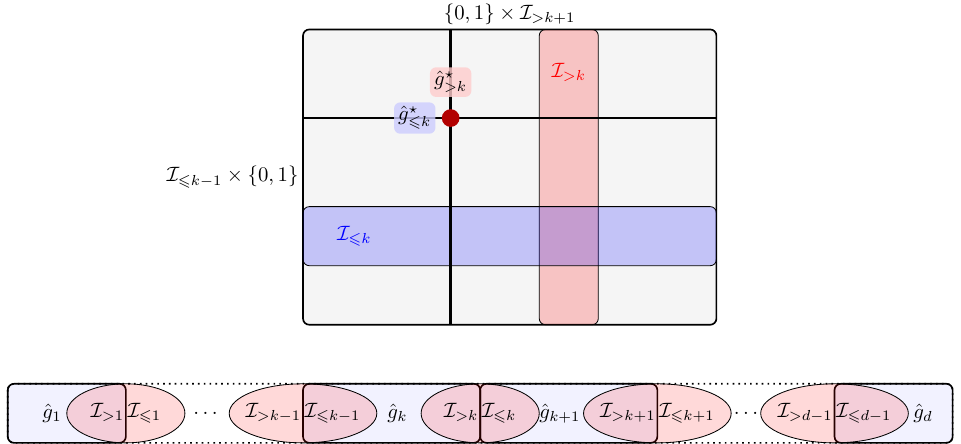}
  \caption{Tensor cross interpolation algorithm visualised.
   Top: one step of tensor cross interpolation updates searches for a new pivot $(\hat g_{\leq k}^\new,\hat g_{>k}^\new)$ in the submatrix $\Like(\I_{\leq k-1}\hat g_k,\hat g_{k+1}\I_{>k+1})$ and adds a new cross to the index set $(\I_{\leq k}, \I_{>k}).$
   Bottom: the process repeats for $k=1,\ldots,d-1$ (left--to--right sweep) and back for $k=d-1,\ldots,1$ (right--to--left sweep).
   } \label{fig:cross}
\end{figure}

Note that in the subtensor interpolation~\eqref{eq:supercore}, the set $\I_{\leq k}$ contains indices, each of which consists of an element from $\I_{\leq k-1}$ and some sample of $\hat g_k$.
Similarly, $\I_{>k}$ samples from the union of $\I_{>k+1}$ and the range of $\hat g_{k+1}$.
This defines \emph{nestedness} relations between the index sets,
\begin{equation}\label{eq:nest}
  \I_{\leq k+1} \subset \I_{\leq k} \times \Binary,\qquad
  \I_{>k} \subset \Binary \times \I_{> k+1}, \qquad k=1,\ldots,d-1,
\end{equation}
which is maintained as long as the indices are only added, never removed.
This allows us to update the interpolations~\eqref{eq:supercore} for any $k=1,\ldots,d-1$, in any order, which enabled efficient parallelisation of this algorithm~\cite{ds-parcross-2020}.
For simplicity in this paper, we update the interpolations sequentially~\cite{sav-qott-2014}, as explained in Alg.~\ref{alg:cross-iter} and shown in Fig.~\ref{fig:cross}.

It remains to iterate the whole process over the TT ranks until an error or complexity stopping criterion is met,
recording the maximal element of $\Like(\g)$ among those seen in the course of iterations.
This is summarized in Algorithm~\ref{alg:cross}.

\begin{algorithm}[t]
  \caption{Full greedy TT cross interpolation and maximization} \label{alg:cross}
  \begin{algorithmic}[1]
  \REQUIRE A procedure to evaluate $\Like(\g)$ for any $\g$, initial index $\g^0 = (\hat g_1^0, \ldots, \hat g_d^0)$, stopping thresholds for error $\delta \ge 0$, TT rank $r_{\max}$ and computational budget $N_{\max}$.
  \STATE Initialise $\I_{\leq k} = \{\hat g_1^0 \ldots \hat g_k^0\}$ and $\I_{>k} = \{\hat g_{k+1}^0 \ldots \hat g_d^0\}$ for $k=1,\ldots,d-1$, $\g^{\max} = \g^0$, $r_0=\cdots=r_d=1$, $\neval=0$.
  \REPEAT
    \STATE Apply Algorithm~\ref{alg:cross-iter} to $\Like(\g)$ starting with $\I_{\leq k}, \I_{>k}$. Find
    \par the new index sets $\I_{\leq k}, \I_{>k}$ for $k=1,\ldots,d-1,$ and
    \par the maximal error $\mathcal{E}$.
    \STATE Update $\g^{\max} = \arg\max \Like(\g)$ over $\g \in \left\{\g^{\max}, (\I_{\leq 1},\I_{>1}), \ldots, (\I_{\leq d-1},\I_{>d-1})\right\}$.
    \STATE Update $r_k:=r_k+1$ for $k=1,\ldots,d-1$.
    \STATE Update $\neval:=\neval + $ number of samples of $\Like(\g)$ drawn in Alg.~\ref{alg:cross-iter}.
  \UNTIL{$\mathcal{E} \le \delta \cdot \Like(\g^{\max})$, $r_k \geq r_{\max}$ $\forall k=1,\ldots,d-1$ or $\neval \geq N_{\max}$.}
  \ENSURE Expanded index sets $(\I_{\leq k}, \I_{>k})$, candidate maximizer $\g^{\max}$.
  \end{algorithmic}
\end{algorithm}

\newpage\clearpage
\section{Tempering and caching of the objective function} \label{sec:tempering}
Efficiency of the TT approximation depends on the decay of singular values of the matricisations $\Like^{\{k\}}$ of the tensor.
In turn, those depend on complexity of the likelihood function.
Although, as we will see in the numerical experiments,
the candidate maximizer produced by Alg.~\ref{alg:cross} can be much more accurate than the approximate tensor itself,
it is still beneficial to be able to vary the complexity of the likelihood for the same data.

One non-intrusive way to achieve this is \emph{tempering}.
Instead of approximating (and maximizing) $\Like(\g)$ directly, we compute the pointwise power of the likelihood, $\Like(\g)^{1/\tau}$.
The parameter $\tau>0$ is similar to temperature in the Boltzmann distribution, hence the name.
To avoid over- and under-flow errors in practical computations, we assemble the tempered likelihood from a sum of logarithms of likelihoods in~\eqref{eq:markov},
\begin{equation}\label{eq:temp}
\Like(\g)^{1/\tau} = \exp\left(\frac{\sum_{k=1}^K \log\Prob(\x_{k-1}\to \x_k | \Grid)}{\tau}\right).
\end{equation}
Note that $\Like(\g)^{1/\tau} \to 1$ as $\tau \to \infty.$
As all entries of the tensor approach $1,$ the whole tensor approaches an array of ones, which has a trivial rank-1 TT decomposition:
\begin{equation}\label{eq:rank1}
 \lim_{\tau\to\infty} \begin{bmatrix}\Like(\g)^{1/\tau} \end{bmatrix}_{\g\in\Binary^d} 
 =
 \begin{bmatrix} 1 \end{bmatrix}_{\g\in\Binary^d} 
 = 
  \begin{pmatrix} 1 \\ 1 \end{pmatrix} 
   \times
  \begin{pmatrix} 1 \\ 1 \end{pmatrix} 
   \times\cdots\times
  \begin{pmatrix} 1 \\ 1 \end{pmatrix}.
\end{equation}
This example shows that by increasing the temperature $\tau$ we reduce the \emph{contrast} of the high--dimensional landscape of the likelihoods, which also reduces the complexity of representation manifested by low TT ranks. 

Increasing the temperature and lowering the contrast generally makes its easier for the tensor cross interpolation algorithm to approximate the tempered high--dimensional likelihood.
However, extremely high temperatures will wipe all contrast and lose all spacial information.
In the  extreme example above, Alg.~\ref{alg:cross} (through Alg.~\ref{alg:mat}) is able to achieve good approximation of~\eqref{eq:rank1} by choosing any interpolation indices, not necessarily those that maximize $\Like(\g).$

In practice, we aim to find an suitable value of $\tau$ that maintains reasonable contrast to allow us to perform optimization, while reducing the TT ranks of the representation and hence the number of computations of the likelihoods required to locate the maximum.
In our numerical examples we find that it is sufficient to find an optimal $\tau$ up to an order of magnitude, which is feasible for trial and error over only a few values.

The TT approximation of 
\(
\Prob(\x_{k-1}\to \x_k | \Grid) = p(\x_k,t_k | \Grid)
\)
notwithstanding, solving many master equations for many data points can take a lot of computing time, e.g. tens to hundreds of seconds per total likelihood evaluation.
Fortunately, the cross structure of samples in Alg.~\ref{alg:cross} contains a lot of repeating indices.
To reduce the computing time, we store a \emph{cache} of all unique indices appearing in Alg.~\ref{alg:cross} together with their (tempered) likelihood values.
Whenever a tensor element is requested at the index which was already considered in the previous iterations, it is taken from the cache instead of evaluating the likelihood from scratch.

\section{Numerical experiments} \label{sec:num}
\begin{figure}[h!t]
  \includegraphics[width=\linewidth]{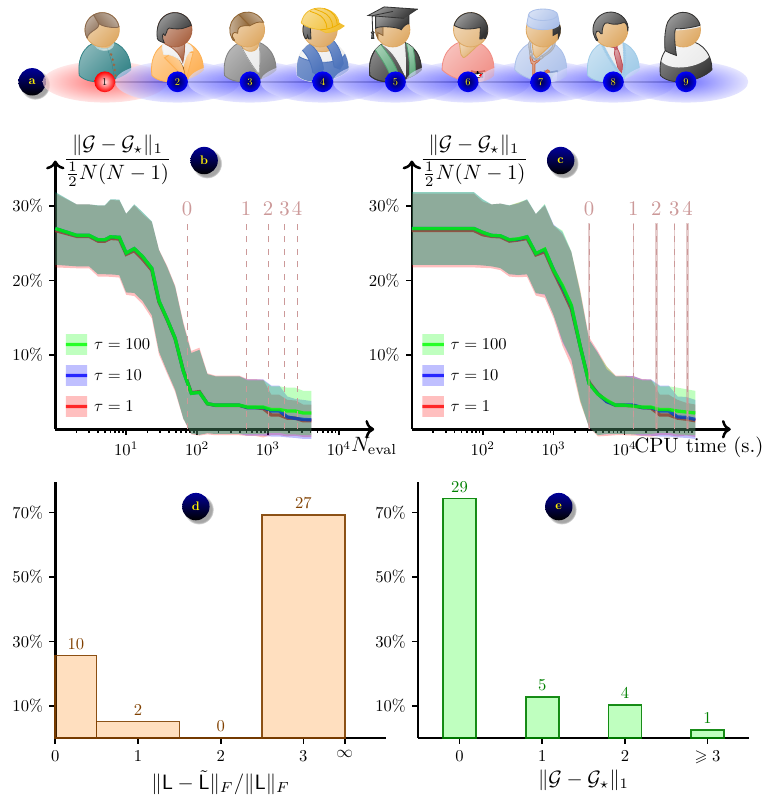}

\caption{Inferring linear chain network with $N=9$ people from $\eps$--SIS epidemic process with $\beta=1,$ $\gamma=0.5$ and $\eps=0.01$:
            (a) the ground truth network $\Grid_\star$ in its initial state;
            (b) and (c) convergence of the network $\Grid$ towards $\Grid_\star$ in the TT cross interpolation algorithm; average (solid lines) $\pm$ one standard deviation (shaded areas) over $N_s=42$ datasets; shown for temperatures $\Temp=1, 10, 100.$
            Vertical dashed lines in (b) and (c) indicate full iterations of the TT cross interpolation algorithm.
            Intervals around the dashed lines in (c) indicate the average time $\pm$ 1 standard deviation.
            (d) distribution of relative errors for the tensor approximation, based on $N_s=39$ datasets;
            (e) distribution of errors for the network inference, based on $N_s=39$ datasets.
                 }
  \label{fig:chain}
\end{figure}

We test the accuracy of the cross interpolation applied for the global optimization of the likelihood by comparing the inferred network $\Grid$ to the `ground truth' network $\Grid_\star,$ which is a linear chain with $N=9$ nodes as shown in Fig.~\ref{fig:chain}(a).
We assume that the initial state 
\(
\x_0=\begin{pmatrix}1 & 0 & \ldots & 0\end{pmatrix}
\)
is the same for all data samples.
We sample random walks of $\eps$--SIS process with parameters $\beta=1,$ $\gamma=0.5$ and $\eps=0.01$ every $\tstep=0.1$ time units for the duration of $\tmax=200$ time units.
This produces $K=2000$ data points $t_k = \tstep k$, $k=1,\ldots,K$, which form the likelihood as shown in~\eqref{eq:like}.

The TT cross interpolation algorithm starts from the index corresponding to the score--based approximate inference as described in~\cite{ds-infer-2024}, and runs for $4$ iterations, achieving the maximal TT rank of $5.$
The master equation~\eqref{eq:cme} is solved on each time interval 
\(
t\in[t_{k-1},t_k]
\)
using the tAMEn algorithm~\cite{d-tamen-2018} with Chebysh\"ev polynomials of degree $12$ in time and a relative stopping and TT truncation threshold $\epsilon_{\text{tAMEn}}=10^{-6}.$

In Fig.~\ref{fig:chain}(b) and (c) we show the convergence of the inferred maximum--likelihood network $\Grid$ towards the ground truth network $\Grid_\star,$ as the TT cross interpolation algorithm progresses.
The error in the network is computed as the number of incorrectly inferred links,
\begin{equation}\label{eq:norm}
 \| \Grid - \Grid_\star \|_1 = \#\{ m,n\in\Nodes, \, m>n : g_{m,n} \neq g_{m,n}^\star \},
\end{equation}
relative to the total number of possible links, $\tfrac12 N(N-1).$
Since the error~\eqref{eq:norm} depends on the particular data in addition to the algorithm, we repeat each experiment (both the data generation and the network inference) $42$ times (the number of available cores on the Intel Xeon Platinum 8168 CPU node used), and plot the average error over those runs, and the interval $1$ standard deviation above and below the average.

Fig.~\ref{fig:chain}(b) shows the convergence with respect to the number of elements the TT cross algorithm draws from the original tensor.
Many of them are repetitive, and hence take almost zero time compared to the likelihood computation at the new samples.
Fig.~\ref{fig:chain}(b) shows the convergence with respect to the actual CPU time.
For this example, about $2/3$ of the samples are retrieved from the cache.
This is a reasonable balance between sparing the computational costs and exploration of the likelihood landscape.

We see that the lowest temperature seems to be most efficient, so potentially one would try to reduce it even further to e.g. $0.1$.
However, for most of the datasets this experiment could not complete, since some likelihood values exceeded the overflow threshold, and the algorithm has stopped.
This indicates that the tempering should be chosen reasonably.
Fortunately, the algorithm is quite insensitive to the particular temperature value within an order of magnitude of the optimal one,
so one needs to try only a few values (such as the orders of 10).

The maximal TT rank of 5 seems too low to approximate a complicated likelihood accurately.
However, the cross algorithm can find a good candidate maximizer even with such small TT ranks, far before a good approximation is reached.
In Fig.~\ref{fig:chain}(d) and (e) we plot the histogram of $\| \Grid - \Grid_\star \|_1$ and the relative mean square error of the whole tensor, $\|\Like - \tilde\Like\|_F / \|\Like\|_F$ for $\tau=1$ over $N_s=39$ experiments (the remaining $3$ experiments did not complete due to overflow in particularly high likelihood values).
For such a high--contrast likelihood we see that the tensor approximation error is actually larger than $1$ for $70\%$ of the data realisations, while $1$ or more incorrect links are found in only $30\%$ of the cases.

\newpage
\section{Conclusion}
In this work, we have shown that the TT cross interpolation algorithm can be used to infer the network structure of the epidemic process from observational data on the states of this process.
The approach is based on the tendency of the rook pivoting in the TT cross interpolation algorithm to find tensor elements close to the global maximum among local maxima of the residual.
It was observed that maximum--volume submatrices tend to contain the maximal element of the matrix~\cite{gt-maxvol-2001},
and here we demonstrated the same effect for the greedy rook pivoting cross interpolation.
The latter requires fewer evaluations of tensor elements after which a consistent approximation or maximization can be produced: the rook pivoting samples only one tensor fiber at a time, while the maximum--volume based methods sample $O(r)$ fibers at once.
A significant further reduction of the computing time ($66\%$ in our example) is achieved by caching of tensor elements, thanks to many cross indices inherited from iteration to iteration in the greedy cross algorithm.

However, the efficiency of maximization can be limited by the restricted nested indices, maintaining the low--rank TT structure.
Moreover, a local maximum of the residual does not have to be the global maximum of the tensor, especially if the tensor contains many nearby elements.
In the latter case we can say that the global maximum has low contrast.
Specifically for the likelihood maximization, we considered tempering as a trade-off between the low--rankness of the tensor of values of the objective function and the contrast of the global maximum.
Intermediate values of the tempering parameter perform best, but it does not have to be very precise,
since the cross interpolation algorithm can reveal a good candidate maximizer even at low TT ranks, that would be insufficient for an accurate approximation of the whole tensor.
It was shown that high-dimensional low-rank canonical tensor decomposition can implicitly denoise the tensor~\cite{Petrov-denoise-2023}.
The TT cross interpolation seems to be recovering most important fibers (and maxima therein) at low TT ranks too.
This paves the way for using the proposed tempered cross interpolation for global maximization of other functions, especially in discrete high--dimensional spaces.
Sharp estimation of TT ranks and optimization performance for tempered functions remains to be a future research. A necessary condition seems to be that the function is strictly uniformly positive, for example, of exponential family, such that the tempered function can be made arbitrarily close to $1.$

\end{document}